\documentstyle[prl,aps,psfig,multicol]{revtex}
\begin{document}
\draft 
\title{ Hidden Quantum Critical Point in a Ferromagnetic Superconductor}
\author{Debanand Sa\cite{e-dsa}}  
\address{Max Planck Institute for the Physics of Complex Systems, 
N\"othnitzer Str. 38, D-01187 Dresden, Germany}  
 

\maketitle

\begin{abstract}
We consider a coexistence phase of both Ferromagnetism and 
superconductivity and solve the self-consistent mean-field equations 
at zero temperature. The superconducting gap is shown to vanish at the 
Stoner point whereas the magnetization doesn't. This indicates that 
the para-Ferro quantum critical point becomes a hidden critical point. 
The effective mass in such a phase gets enhanced whereas the spin wave 
stiffness is reduced as compared to the pure FM phase. The spin wave 
stiffness remains finite even at the para-Ferro quantum critical point. 

\end{abstract}
\begin{multicols}{2}  
\narrowtext 
The study of itinerant Ferromagnetic materials are becoming more and 
more important due to the role of strong electronic correlations and 
the appearance of many exotic phases such as non-Fermi liquid (NFL), 
superconductivity (SC) etc. near the para-Ferro quantum critical point 
(QCP). In earlier times, Ferromagnetism (FM) was believed to suppress  
SC but the recent discovery of SC \cite{sax00} below 1 K in the 
pressure range of 1 to 1.6 GPa in a high purity single crystal of UGe$_2$ 
has ruled out the above possibility. This rather suggests that FM and SC 
could be cooperative. Also, there are other materials such as ZrZn$_2$ 
\cite{pfl01} and URhGe \cite{aok01}, where the coexistence of FM and SC 
has been found. SC phase in all the above mentioned materials is completely 
covered within the FM phase and disappears in the paramagnetic (PM) region. 

The standard way to look for a coexistence phase of FM and SC 
theoretically is to introduce two kinds of fermions. FM could be caused  
by local f-electrons  whereas SC, by itinerant ones. But in the above 
materials such as, UGe$_2$ and URhGe, both the roles 
were played by the same Uranium 5-f electrons which are itinerant and  
strongly correlated. Thus, it would be interesting to study microscopically 
a model where the coexistence of both FM and SC can be described by only one 
kind of electrons. Such a model study has recently 
been initiated by Karchev et. al. \cite{kar01}. Of course, this model 
is confined to singlet SC which is unlikely to occur inside a FM.  

In this letter, we study the coexistence phase of both FM and SC and 
look for the consequences. We solve the zero temperature self-consistent 
mean-field equations. It is shown that the SC gap vanishes at the 
para-Ferro QCP in such a model whereas the magnetization doesn't. This 
suggests that the SC pairing induces a small but finite magnetization 
which doesn't vanish even at the Stoner threshold. This is an indication 
of the para-Ferro QCP becomming a hidden QCP. We also computed the 
effective mass as well as the spin wave stiffness in the coexistence 
phase. The later eventhough reduced, is nonzero at the para-Ferro QCP. 

We presume that the relevant magnetic behaviour of the system is 
adequately described by Stoner RPA-mean-field theory \cite{bri68} 
and do not question on its stability. This can be obtained from 
Hubbard Hamiltonian. In order to get SC pairing, we add a reduced 
BCS Hamiltonian to it. We would like to mention here that the pairing 
Hamiltonian is not due to the FM spin fluctuation rather it may be due to 
some other means. Thus, to describe a coexistence phase of both 
FM and SC, one can start with a minimal effective Hamiltonian,

\begin{equation} 
H=\sum_{k,\sigma} \epsilon_k c_{k\sigma}^\dagger c_{k\sigma}  
+U\sum_i n_{i\uparrow} n_{i\downarrow} 
-V\sum_{k,k'} c_{k\uparrow}^\dagger c_{-k\downarrow}^\dagger 
c_{-k'\downarrow} c_{k'\uparrow},  
\end{equation} 
 
\noindent where $U$ and $V$ are respectively the on-site Hubbard 
interaction energy and the reduced BCS pairing energy.   
$n_{k\sigma}=c_{k\sigma}^\dagger c_{k\sigma}$ is the electron density 
and $c_{k\sigma}^\dagger (c_{k\sigma})$ are the standard electron 
creation (annihilation) operator with wave vector $k$ and spin projection 
$\sigma$. In order to obtain a coexistence phase of both FM and SC, we 
can perform a mean-field theory by defining the averages, 
$2\Delta_F=U(<n_{\downarrow}>-<n_{\uparrow}>)$, and 
$\Delta=V\sum_k <c_{k\uparrow}^\dagger c_{-k\downarrow}^\dagger>$.  
Here $2\Delta_F$ and $\Delta$ are respectively the FM and the 
SC order parameter in the coexistence phase. At this stage, one can 
diagonalize the above Hamiltonian through the standard Bogoliubov 
transformation and the new energy dispersions are obtained as, 

\begin{equation} 
E_k^{\alpha}=\Delta_F + \sqrt{(\epsilon_k-\mu)^2 +|\Delta|^2}, 
\label{DeltaF}   
\end{equation} 

\begin{equation} 
E_k^{\beta}=\Delta_F - \sqrt{(\epsilon_k-\mu)^2 +|\Delta|^2}.  
\label{Delta} 
\end{equation} 

\noindent  The subscript $\alpha$ and $\beta$ above denote the two 
different Bogoliubov Fermions in the coexistence phase.  
The self-consistent mean-field equations are derived as, 

\begin{equation} 
2\Delta_F = \lambda \int d\epsilon  
(1-n_k^{\beta} - n_k^{\alpha}), \label{lambda}     
\end{equation} 

\begin{equation} 
|\Delta| = g \int d\epsilon {{|\Delta|}\over {2 E}}
(n_k^{\beta} - n_k^{\alpha}), \label{g}    
\end{equation} 
 
\noindent where $n_k^{\beta,\alpha}$ are the momentum distribution 
for the corresponding Bogoliubov Fermions and $\lambda=U\rho(0)$, 
$g=V \rho(0)$, $E_k=\sqrt{(\epsilon_k-\mu)^2 +|\Delta|^2}$, 
$\rho(0)$, being the density of state at the Fermi level in the PM phase. 
 
It is obvious from equations \ (\ref{DeltaF}) and (\ref{Delta}) 
that for $\Delta_F > 0$, $E_k^{\alpha} > 0$ for all $k$ and thus, at 
zero temperature, $n_k^{\alpha}=0$. On the other hand, for $E_k^{\beta}$, 
there are two possibilities, (i) $E_k^{\beta} > 0$ and 
(ii) $E_k^{\beta} < 0$.  In case (ii) $n_k^{\beta}=1$ for all $k$. 
Substituting this in equation \ (\ref{lambda}) yields  $\Delta_F = 0$. 
Thus, the only solution for equation \ (\ref{lambda}) which allows  
nonzero $\Delta_F$ in order to get a coexistence phase is  $E_k^{\beta} > 0$. 
The dispersion of the $\beta$-Fermion becomes positive only in the 
energy interval $\epsilon_F^{-} < \epsilon_k < \epsilon_F^{+}$, where 
$\epsilon_F^{\pm}$ are the solutions of the equation $E_k^{\beta}=0$, 
which is given as, $\epsilon_F^{\pm}=\epsilon_F\pm \sqrt{\Delta_F^2 
-|\Delta|^2}$. It should be noted here that $\epsilon_F^{\pm}$ are the 
new Fermi energies in the coexistence phase. However, to get a nonzero 
$\Delta$ from equation \ (\ref{g}), one can have  $E_k^{\beta} < 0$, 
which corresponds to the case $\epsilon_F^{-} > \epsilon_k > 
\epsilon_F^{+}$. Thus, the above self-consistent mean-field equations \ 
(\ref{lambda}) and (\ref{g}) at $T=0$ take the form,  

\begin{equation} 
2\Delta_F = \lambda \int_{\epsilon_F^{-}}^{\epsilon_F^{+}} 
d\epsilon, 
\label{lambda1}     
\end{equation} 

\begin{equation} 
|\Delta| = g |\Delta|  (\int_{-W/2}^{W/2} 
-\int_{\epsilon_F^{-}}^{\epsilon_F^{+}})  
d\epsilon {{1}\over {\sqrt{(\epsilon-\mu)^2+|\Delta|^2}}},  
\label{g1}    
\end{equation} 
 
\noindent $W$ being the band width. 
Equation \ (\ref{lambda1}) can be solved analytically  
and the FM order parameter is obtained as, 

\begin{equation} 
\Delta_F = {{\lambda}\over {\sqrt{\lambda^2 -1}}}|\Delta|. \label{gap}    
\end{equation} 

\noindent Now, the SC order parameter from equation \ (\ref{g1}) can be 
computed by assuming the standard procedure of integration in a shell 
$(\Lambda/2)$ around $\epsilon_F$. In this approximation, 
equation \ (\ref{g1}) reduces to, 

\begin{equation} 
{1\over {g}} = (\int_{\epsilon_F -\Lambda/2}^{\epsilon_F +\Lambda/2} 
-\int_{\epsilon_F^{-}}^{\epsilon_F^{+}})  
d\epsilon {{1}\over {\sqrt{(\epsilon-\mu)^2+|\Delta|^2}}}.    
\end{equation} 

\noindent where $\epsilon_F +\Lambda/2 > \epsilon_F^{+}$ and 
$\epsilon_F -\Lambda/2 < \epsilon_F^{-}$. Performing the integration 
and substituting $\Delta_F$ from equation  \ (\ref{gap}), one obtains,  

\begin{equation} 
|\Delta| = \sqrt{{\lambda -1}\over{\lambda +1}} 
\Lambda   e^{-1/g}. \label{gap0}   
\end{equation} 

\noindent Thus, $\Delta_F$ can be calculated from equation \ (\ref{gap}) 
as,  

\begin{equation} 
\Delta_F = {{\lambda}\over{\lambda +1}} \Lambda e^{-1/g}. \label{gap1}
\end{equation} 

\noindent Furthermore, putting the above values of $\Delta_F$ and $\Delta$ 
in the expression for $\epsilon_F^{\pm}$, one obtains, 

\begin{equation} 
\epsilon_F^{\pm} = \epsilon_F \pm {1\over{\lambda +1}} \Lambda e^{-1/g}. 
\label{fermi} 
\end{equation} 

\noindent The above equations \ (\ref{gap0}), (\ref{gap1}) and (\ref{fermi}) 
are of crucial importance in the present manuscript. It is clear from these 
equations that the SC gap $\Delta$ as well as the uniform magnetization 
($\propto \Delta_F$) decrease as one approaches the Stoner threshold 
($\lambda =1$). $\Delta$ vanishes exactly at $\lambda =1$ whereas 
$\Delta_F$ doesn't. This is an indication that the SC pairing induces 
spontaneous magnetization in the system which 
does not vanish at Stoner threshold (In principle $\Delta_F =0 $ at 
Stoner threshold). Thus, the para-Ferro QCP becomes a hidden one due 
to the presence of SC pairing. Furthermore, the new Fermi energy  
$\epsilon_F^{\pm}$ moves away more and more from $\epsilon_F$, as one 
approaches the Stoner point. It becomes exactly equal to the Fermi 
energy of the Stoner FM ($\epsilon_F^{\pm}=\epsilon_F \pm \Delta_F$) 
at the Stoner threshold. 
 
Next, let us consider the distribution functions $n_k^{\uparrow}$ and 
$n_k^{\downarrow}$ for the spin up and spin down quasi particles in terms 
of the Bogoliubov Fermions. These are already discussed in an earlier 
literature \cite{kar01} and for completeness we can write them as,   

\begin{eqnarray} 
n_k^{\uparrow}=u_k^2 n_k^{\alpha} + v_k^2 n_k^{\beta} \nonumber \\ 
=v_k^2 [\theta(k_F^{-}-k)+ \theta(k-k_F^{+})],      
\end{eqnarray} 

\begin{eqnarray} 
n_k^{\downarrow}=1- u_k^2 n_k^{\beta} - v_k^2 n_k^{\alpha} \nonumber \\
=n_k^{\uparrow} +[\theta(k_F^{+}-k)- \theta(k_F^{-}-k)],  
\end{eqnarray} 

\noindent where $u_k^2$ and $v_k^2$ are the coherence factors involved 
in the Bogoliubov transformation which have the standard form in any  
mean-field theory. As already discussed before, at $T=0$, $n_k^{\alpha}=0$ 
and $ n_k^{\beta}=\theta(k_F^{-}-k)+ \theta(k-k_F^{+})$. 
$k_F^{\pm}$ are the wave vectors corresponding to the new Fermi  
energy $\epsilon_F^{\pm}$. It should be noted at this point that, in 
a standard SC theory, a gap appears around the Fermi surface, but 
in the present case, Fermi surfaces appear for the Bogoliubov Fermion 
$\beta$ in the coexistence phase which is completely unexpected. This 
could be due to the fact that the itinerant FM had already have 
the Fermi surfaces which still persist in the coexistence phase. 
Therefore, the existence of two Fermi surfaces is 
a generic property of the coexistence phase of both FM and 
SC since it is caused by the same quasi particles in the system. These 
Fermi surfaces are already reflected in the spin up and down momentum 
distribution functions and will lead to different properties in the 
system as compared to a  standard mean-field theory. The single  
particle density of states which appears in almost all the properties 
of the system turns out to be, 

\begin{equation} 
N(0)={{\rho(0)(\epsilon_F^{+}+\epsilon_F^{-})\Delta_F}\over {2\epsilon_F 
\sqrt{\Delta_F^2 -|\Delta|^2}}}
=N^{+}(0) + N^{-}(0), \label{dos} 
\end{equation} 

\noindent  where $ N^{+}(0)$ and  $N^{-}(0)$ are respectively the density 
of states on the two Fermi surfaces $\epsilon_F^{+}$ and $\epsilon_F^{-}$ 
of the Bogoliubov Fermion $\beta$. For finite $\Delta_F$, the density of 
states increases with $\lambda$, as opposed to the case of a standard FM 
metal. The presence of Fermi surfaces together with the enhanced density 
of states at the Fermi level have 
important consequences in the thermodynamic properties of the 
system. The specific heat capacity, for example, at low temperature 
shows linear temperature dependence ($C_v (T)=\gamma T$) as opposed 
to the activated behaviour. This can again be understood in terms of  
the presence of Fermi surfaces of the $\beta$-Fermion. Moreover, 
the $\gamma$-coefficient in the specific heat which depends on 
$N(0)$ also gets enhanced due to increase in the density of states. 

The increase in the single particle density of states can be 
understood in the following way: One can investigate the changes 
in the energy dispersion of the $\beta$-Fermion due to the appearence 
of the new Fermi energy in the coexistence phase. Substituting the 
expression for $\epsilon_F^{\pm}$ in $E_k^{\beta}$ and approximating 
${{\epsilon_k -\epsilon_F} \over{\Delta_F}}\ll 1$, one can obtain the 
energy dispersion for the $\beta$-fermion as, 

\begin{equation} 
E_k^{\beta} \approx \pm {{\sqrt{\Delta_F^2 - \Delta^2}}\over{\Delta_F}} 
(\epsilon_k -\epsilon_F), \label{dis} 
\end{equation}

\noindent which is just the renormalized free Fermion dispersion.  
The renormalization factor ${{\sqrt{\Delta_F^2 - \Delta^2}} 
\over{\Delta_F}}$ enters not only in the energy dispersion but also 
in the density of states which is obvious from equation \ (\ref{dos}) 
and \ (\ref{dis}). Thus, the enhancement in the density 
of states at the Fermi level can be thought to be due to 
the reduction in the band width. This can also cause an increase in 
the effective mass ($m^*= {{m \Delta_F}\over{\sqrt{\Delta_F^2- 
\Delta^2}}}$) similar to that of density of states. However, the 
enhancement in the density of states/effective mass or the reduction 
in the $\beta$-Fermion band becomes prominent when one moves away 
from the para-Ferro QCP. This is due to the fact that the  
renormalization factor becomes unity at the Stoner point.  

Let us now consider the effect of induced magnetization due to SC pairing 
in the spin wave dispersion. This can be achieved by analyzing the RPA 
transverse susceptibility \cite{izu63} in the coexistence phase, which 
is given as, 

\begin{equation} 
\chi_{RPA}^{+-}(q,\omega)= {{\chi_0^{+-}(q,\omega)}\over{1-U 
\chi_0^{+-}(q,\omega)}}, 
\end{equation} 

\noindent where $\chi_0^{+-}(q,\omega)$ is the unperturbed 
transverse susceptibility in the coexistence phase. 
Using the expressions for the Bogoliubov coherence 
factors, it can be computed as, 

\begin{eqnarray} 
\chi_0^{+-}(q, \omega)= 
{1\over{4}}\sum_k (1-{{\epsilon_k \epsilon_{k+q} 
+\Delta^2}\over{E_k E_{k+q}}})\nonumber \\ 
({1\over{\omega +2\Delta_F +E_{k+q} +E_k}} 
+{1\over{\omega +2\Delta_F -E_{k+q} -E_k}})\nonumber \\ 
+{1\over{4}}\sum_k (1+{{\epsilon_k \epsilon_{k+q} 
+\Delta^2}\over{E_k E_{k+q}}})\nonumber \\ 
({1\over{\omega +2\Delta_F +E_{k+q} -E_k}} 
+{1\over{\omega +2\Delta_F -E_{k+q} +E_k}}).   
\end{eqnarray}

\noindent Spin wave dispersion can be obtained from the divergence of 
$\chi_{RPA}^{+-}(q,\omega)$, i. e., from the solutions of the equation 
$1-U\chi_0^{+-}(q,\omega)=0$. Expanding $\chi_0^{+-}(q,\omega)$ for 
small $q$ and $\omega$ and for ${{\omega}\over{2\Delta_F}}\ll 1$, the 
spin wave dispersion turns out to be, 

\begin{equation} 
\omega= D q^2,   
\end{equation}   

\noindent where the spin wave stiffness D is computed as,  
$D={{1}\over{18}}{{\Delta_F 
\sqrt{\Delta_F^2 -\Delta^2}}\over{\epsilon_F^2 m}}$,   
$m$ being the bare electron mass. The spin wave stiffness is reduced 
compared to that in the pure FM phase and becomes finite even at the 
Stoner critical point. This is due to the fact that the induced   
magnetization caused by SC pairing in the coexistence phase remains 
finite at the Stoner critical point.

Another important feature of the coexistence phase is the appearance 
of Fermi surfaces in the system. The consequence of this is the presence 
of paramagnons which describe the longitudinal spin fluctuations 
\cite{izu63}. They not only survive in the FM metallic phase but also 
in the coexistence phase of both FM and SC. The propagator for the 
longitudinal spin fluctuations is given as, 

\begin{equation} 
\chi_l(q,\omega)= {{1}\over {\eta + b q^2 + {{ic|\omega|}\over {q}}}}, 
\end{equation} 

\noindent where b and c are constants depending on the parameters 
in the system and $\eta$, which is the inverse of the static 
susceptibility is given by, 

\begin{equation} 
\eta= 1-UN(0).
\end{equation} 

\noindent  As we have already mentioned earlier, the density of states 
equation \ (\ref{dos}), increases with $\lambda$ which makes the  
inverse of the static susceptibility $\eta$ to vanish even if for small 
$\Delta_F$. This is quite different from that of weak FM metals where 
$\eta$ becomes zero at zero magnetization. Thus, the finite value of 
induced magnetization makes the Stoner QCP hidden.  

The results obtained for the coexistence phase in the present manuscript 
is described only in the mean-field level which becomes a starting point 
for going beyond it. Since the appearance of induced magnetization in the 
coexistence phase makes the para-Ferro QCP a hidden one, it would be 
important to investigate the role of quantum fluctuations on it which is 
left for future study \cite{deb02}. The conclusion that the para-Ferro 
QCP becomes a hidden one has also been pointed out recently in case of the 
coexistence of FM and spin triplet SC \cite{spa02}. Thus, one can conclude 
that the hidden QCP might be a generic property of the coexistence phase 
where both the spin rotational and the gauge symmetry are broken and is 
independent of the symmetry of the SC order parameter.  

However, in the present work, the 
SC QCP is dressed in the sense that SC can occur at zero magnetization.  
This is due to the fact that the SC pairing is caused not by spin  
fluctuations rather by some other means such as phonons. This could be 
contrasted with the standard spin fluctuation theory in an itinerant 
FM \cite{fay79} where the QCP is naked. In the later case, the FM-SC 
transition temperature vanishes at the QCP. From the above scenarios,   
it might be possible to differentiate whether the SC in a FM is due to 
spin fluctuations or by some other means. The materials about which   
we mentioned at the beginning of the present manuscript fall into 
the second category where both SC and FM transition temperature vanish 
at the QCP. Thus, the SC mechanism in these materials might be thought 
to be due to spin fluctuations.   

In conclusion, we briefly outline our findings. We consider a  
possible coexistence phase of both FM and SC. We solve the 
self-consistent mean-field equations for the uniform magnetization 
as well as the SC order parameter. It has been shown that both the order   
parameter decrease as one approaches the Stoner critical point. The SC gap 
vanishes exactly for $\lambda=1$ but on the contrary, the uniform  
magnetization doesn't. This shows that the SC pairing induces a finite 
nonzero magnetization in the coexistence phase which washes out the Stoner  
QCP and makes it hidden. 
Moreover, we computed the effective mass as well as the 
spin wave dispersion in the coexistence phase. The former is enhanced 
but the later gets reduced and remains finite even at the Stoner  
threshold. Furthermore, the Bogoliubov Fermions in the coexistence phase  
retains the Fermi surfaces, which gets reflected in the thermodynamic 
properties of the system. In particular, the specific heat capacity has 
linear temperature dependence as in the standard itinerant FM, but 
the $\gamma$-coefficient increases anomalously for a small magnetization. 
This is due to the fact that the single particle density of state in the 
coexistence phase gets enhanced.   

\smallskip   
The author would like to thank Amit Dutta for carefully reading the 
manuscript.  






\end{multicols}  
\end{document}